\documentstyle[prl,aps,psfig]{revtex}
\begin{document}
\draft
\twocolumn
\title{Infinite magnetoresistance of magnetic multilayers}

\author{R. Seviour$^*$, S. Sanvito$^{\dagger*}$, C.J. Lambert$^*$
 and J.H.Jefferson$^{\dagger}$}
\address{$^*$ School of Physics and Chemistry,
Lancaster University, Lancaster LA1 4YB, U.K.\\
$^{\dagger}$Defence Evaluation and Research Agency,  EOMC,
Malvern, Worcs. WR14 3PS UK}
\date{\today}
\maketitle

\begin{abstract}
 We examine transport properties of a magnetic 
superlattice with current perpendicular to the planes. In the 
limit that the phase-breaking and spin flip scattering lengths 
are greater than the system size, a multiple-scattering approach 
is used to calculate the 4-probe conductance. We show that by tuning 
the strength of tunnel barriers placed between the current and 
voltage probes giant magnetoresistance ratios of arbitrary strength 
and size are achievable.
\end{abstract}

\smallskip


 During the past few years advances in fabrication technology have enabled the 
production of precisely engineered mesoscopic superlattice (SL) structures, 
consisting of alternating magnetic and non-magnetic metallic layers, with well 
defined dimensions and interfaces. Exchange coupling of the magnetic layers
through the non-magnetic material gives rise to antiferromagnetic (AF) alignment
of adjacent magnetic layers. When such AF alignment is broken by applying a 
large magnetic field, a global ferromagnetic (F) configuration of the multilayer
is achieved, and the resistance drops drastically \cite{bab,brot,r1,r2}. 
Such functional magnetic materials are the focus of substantial research 
both from a technological and fundamental viewpoint, having a wide range of 
applications in magneto-electronics \cite{pt}, such as read/write heads
for high density magnetic storage systems, and in the miniaturization
of magnetic field sensors, such as solid state compasses \cite{siem}.
All these applications require materials/geometries capable of giving a
large signal (i.e. a large drop in resistance), and high sensitivity
(so that only a small magnetic field required). Conventional materials, 
such as the 3$d$ transition metal multilayers, show large GMR, but high 
magnetic fields are necessary to overcome the exchange coupling, making such
systems unsuitable for most applications. For inhomogeneous 
multilayers based on polycrystalline alloys \cite{siem} much smaller 
fields are required, but show quite small drops in resistance. 
The half-metals such as CrO$_2$, in principle can guarantee an infinite 
GMR \cite{alex1,alex2}, but to date a GMR of only 50\% has been measured 
in CrO$_2$ powders at low temperature \cite{coey}.

 In  this letter we argue that attention should also focus on the problem of 
optimising the scattering properties of a given structure and 
on the role of external probes. By tunning the scattering potential in a 
4 - probe geometry we shall demonstrate a significant GMR enhancement can be 
achieved.

 Transport properties of GMR structures are typically studied, when the current 
flows either in the plane of the layers (CIP) or perpendicular to the
 plane of the layers (CPP). In the CIP configuration the dimensions are 
macroscopic and 4 probe measurement techniques are used \cite{park}. In 
contrast CPP measurements are usually carried out with 2 probes, using 
sophisticated techniques to measure the small resistances involved 
($\sim 10^{-8}\Omega$). The CPP configuration is particularly interesting 
because it produces a larger GMR signal, with changes in resistance up to 
$150$\% \cite{r3,r4}. Remarkably however very little research has been 
conducted using 4 probe measurements on CPP geometries \cite{phil}, 
despite the fact that the quantum nature of transport in such structures 
gives rise to unexpected phenomena, as detected in anisotropic 
magnetoresistance (AMR) systems of Co/Ni multilayers \cite{r5,r6}.

 In this letter we consider a 4 probe CPP structure, and demonstrate that 
due to an inherent instability of 4 probe conductance measurements, an 
infinite magnetoresistance (IMR) is achievable.
We use a general scattering approach to dc transport,
developed to describe phase-coherent transport in dirty
mesoscopic structures, and based on the fundamental current-voltage 
relations derived in \cite{rbut}. For simplicity in this Letter, 
we focus on the 
conductance of the structure shown in Fig.1(a),
which comprises a superlattice, with alternating normal 
and magnetic metal layers, in contact with four normal 
reservoirs at voltages $v_j  (j=1,\dots ,4$). The leads connecting the 
structure to  reservoirs 1 and 2 carry a current $I$, whereas 
 leads 3 and 4 carry no current and hence form the voltage probes. The 
current/voltage leads are separated by an 
insulating barrier.

 
 In what follows, we consider the zero-temperature, zero-bias limit 
in which the phase-breaking and spin-flip scattering lengths are 
 greater than the dimensions of the system. In this limit, transport properties 
depend not only on the electronic structure of the magnetic multilayers, but 
also on the contacts with external reservoirs. This feature is embodied in a 
fundamental current - voltage relation due to B\"{u}ttiker \cite{rbut},

\begin{equation} 
I_i=\sum_{j=1}^4G_{ij} v_j \;{.}
\label{eq1}
\end{equation}


which relates the current $ I_i$ 
from a normal reservoir $i$ to the reservoir voltages $(v_j)$. 
The coefficients $G_{ij}$ satisfy $\sum_{j=1}^4G_{ij}
=\sum_{i=1}^4G_{ij} = 0 $. In units of
$2e^2/h$ \cite{rbut}, $G_{ii}=N_i-R_i$ and $G_{ij\ne i}=-T_{ij}$,
where  $T_{ij}$ is the transmission coefficient from
probe $j$ to probe $i$, $R_i$ is the reflection
coefficient in probe $i$ and $N_i$ is the number of open scattering
channels in lead $i$. As indicated in Fig. 1b, Eq.(\ref{eq1}) can be 
viewed as an equivalent circuit representation of the phase coherent 
structure of Fig.1a. However the elements $G_{ij}$ are correlated 
functionals of the scattering potential generated by contact 
with external leads, geometry and disorder, and therefore cannot be 
varied independently. Furthermore, as noted in \cite{rbut}
$G_{ij} (H) = G_{ji} (H^*)$, where $H$ is the Hamiltonian of the system
and therefore in the presence of a magnetic field,  $G_{ij}\neq G_{ji}$.

Setting $I_1 = -I_2 = I, I_3=I_4=0$ and solving 
Eq.(\ref{eq1}) for the conductance yields,

\begin{equation} 
G= \frac{h}{2e^{2}} \frac{I}{(V_{3} - V_{4})} = \frac{d}{(G_{42}G_{31} -
 G_{41}G_{32} )} \;{,}
\label{eq2}
\end{equation}

where $d \geq 0$ is the determinant of the $3{\rm x}3$ matrix obtained by 
removing the third row and column from the $G_{ij}$ matrix of Eq.(\ref{eq1}). 
In what follows we compute the GMR ratio (${\tilde G}$) as \cite{r2}

\begin{equation}
{\tilde G} = \frac{ G_{\downarrow} + G_{\uparrow} -2 G_{\downarrow\uparrow}}{
 2 G_{\downarrow\uparrow}} \;{,}
\label{eq4}
\end{equation}

 where $G_{\downarrow}$  $(G_{\uparrow})$ is the  
conductance for the minority (majority) spin in the
F configuration, and  $G_{\downarrow\uparrow}$ is the conductance 
for both spins in the AF configuration.


In the F and AF configurations, the various transmission 
and reflection coefficients for the separate spins can be computed by 
solving the
Schr\"odinger equation on a nearest neighbor tight-binding lattice of
sites. In simple quasi one dimensional gemoetries involving only two probes, 
by combining an efficient recursive Green's function approach with a 
material-specific s-p-d tight-binding Hamiltonian one can accurately predict 
the GMR ratio for a range of materials and layer thickneses \cite{noi}.
 For more complex geometries involving several probes, this is beyond the 
capabilities of currently available computing resources and therefore to 
demonstrate a generic enhancement of the GMR ratio, we analyze a tight 
binding model 
with a single degree of freedom per spin on each lattice site, as introduced in 
\cite{jap1,jap2}.
 Each lattice site is labeled by an index $i$ and possesses 
a spin degree of freedom $\psi_i^{\sigma}$, and the  corresponding 
Schr\"odinger equation has the form

\begin{equation}
E\psi_i^{\sigma}
=\epsilon_i^{\sigma} \psi_{i}^{\sigma}
-\sum_{j} (\gamma_{ij} \psi_{i}^{\sigma})\;{,}
\label{eq5}
\end{equation}

where $j$ sums over all neighbors of $i$. The nearest neighbor hopping 
elements $\gamma_{ij}$  
fix the band-width and $\epsilon_i^{\sigma}$ determines the
band-filling. The separate spin fluids are assigned a 
spin-dependent on-site energy $\epsilon_i^{\sigma}$. The parameters used in 
the present 
calculations are as follows: in the
leads and  non-magnetic regions of the SL 
$\epsilon_i^{\sigma}=1$ and $\gamma_{ij}=1$; in the magnetic layers
$\gamma_{ij}=2$ and $\epsilon_i^{\sigma}=1.45$ ($3.7$) 
for spin up (down). Following \cite{noi} the hopping elements $\gamma_{ij}$ 
joining magnetic-normal sites are chosen to be the geometric mean of the 
hopping elements in the magnetic/normal regions.

The dc conductance (\ref{eq2}) is determined by the multi-channel quantum 
mechanical scattering matrix, which in turn can be found from the 
Green's function for a given structure. The numerical approach involves 
attaching the scattering region to crystalline semi-infinite leads, in which 
the Green's functions are superpositions of plane waves, and then using 
Dyson's equation to 
determine the total Green's function \cite{noi}. Effectively this procedure requires 
the inversion of Hamiltonian describing the scatterer, which for the structure 
analyzed here is a $6080X6080$ matrix per spin degree of freedom. As we are only 
interested in the Green's functions on the surface of the structure the 
inversion is carried out after the Hamiltonian has been renormalised in 
real space using an exact decimation technique \cite{ro1}. 

As shown in Fig.1a the structure analyzed consists 
of four, crystalline semi-infinite leads each 40 
sites wide. The superlattice consists of 
5 non-magnetic layers alternating with 4 magnetic layers, each 8 sites 
wide and 40 sites long. The leads are separated by barriers, 2 sites 
wide and of strength $U$. These barriers change the on-site energy  
$\epsilon_i^{\sigma}$ in Eq.(\ref{eq5}) to $\epsilon_i^{\sigma}+U$. Hence, 
varying the magnitude of $U$ alters the transmission 
and reflection coefficients of the current/voltage leads. For a 
barrier of width 2 sites,  
fig.2 shows the  dependence on $U$ of the transmission 
coefficient per open channel, which as expected, decays exponentially with $U$ 
for large $U$.

 Fig.3 shows the dependence of $G_{\uparrow}$,$G_{\uparrow\downarrow}$ 
and $G_{\downarrow}$ on barrier strength. We see that 
$G_{\uparrow\downarrow}$ and $G_{\downarrow}$ increase smoothly as $U$ is 
increased,  whereas $G_{\uparrow}$
 diverges for $U \approx 2.2$, and then returns to a negative constant. 
This divergence occurs as the transmission coefficient from lead 
4 to lead 1 approaches the transmission coefficient from lead 4 to lead 2. 
 
 To gain physical insight into the origin of this divergence,  
we return to the equivalent circuit of fig.1b and consider the case of a 
symmetric structure, 
where $G_{13}$ = $G_{24}$,
$G_{12}=G_{34}$, $G_{14}=G_{23}$ and Eq.(\ref{eq2}) reduces to

\begin{equation}
G=\frac{2(G_{43}+G_{42})(G_{43}+G_{41})}{G_{42}-G_{41}} \;{.}
\label{eq6}
\end{equation}

 From this expression we see that if 
$G_{42}<G_{41}$ then the 4 probe conductance will be negative 
and when $G_{42}=G_{41}$, the conductance is infinite. Of course 
since the quantities $G_{ij}$ are determined by the quantum 
mechanical scattering properties of the structure in figure 1, there is 
no guarantee that such a condition is achievable. The fact that we are able to 
demonstrate that this can occur in a realisable structure is a key result of this 
Letter.

 Fig.4 shows the dependence  of the GMR ratio on barrier strength, 
calculated using Eq.(\ref{eq4}). Due to the behavior of the 
majority spin state ($G_{\uparrow}$), the  GMR ratio 
 increases with increasing $U$, 
diverges for $U \approx 2.2$ giving an IMR. Further increasing U 
causes the GMR ratio to return from -$\infty$ to a 
constant negative value, of around $-110\%$.
We note that such negative GMR is not due to an intrinsic
scattering asymmetry of the materials as measured in Ref.\cite{fert}, 
but is due to the peculiar spin-dependent behavior 
of the 4 probe measurement. It is important to point out 
that the system we described is in principle tunable,  
for example by applying a gate electrode giving rise to the 
barrier potential U. Hence by changing the 
external gate potential, arbitrary GMR ratios may be obtained, 
using exactly the same principle as a Wheatstone bridge, 
despite the fact that the system described is fully quantum mechanical.

In conclusion, we have shown for the first time that due to an inherent 
instability in the 4-probe conductance of mesoscopic structures 
an infinite GMR ratio is realizable.  To-date a great deal of effort 
has been aimed at 
optimizing the materials used in GMR devices. The above results suggest 
that geometry and external gating may be equally crucial in the race to 
obtain inexpensive magnetoresistive devices.
 This effect is a generic feature of the 4 probe geometry,
and the use of a simple single-band tight-binding
model is sufficient to demonstrate the possibility of IMR. 
Nevertheless more sophisticated material-specific
tight-binding models \cite{oxo,noi}, or {\it ab-initio} calculations 
\cite{baur}, will be needed to provide quantitative insights into real device 
characteristics.

Acknowledgments

 The authors wish to thank P.wright, A. Volkov, R. Raimiondi and V. Falko for 
detailed discusions. This work is funded by the EPSRC and the E.U. TMR 
programme.


\begin{figure}
\centerline{\psfig{figure=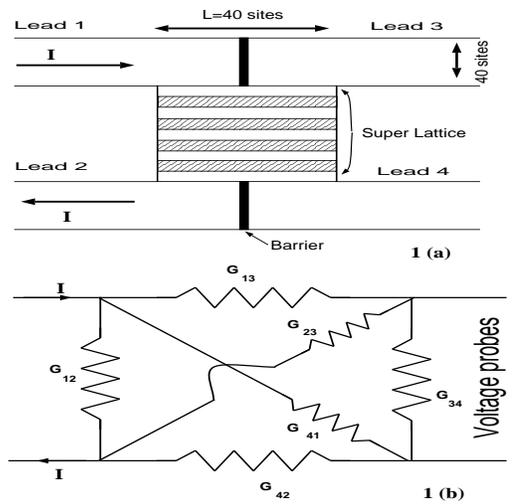,width=7cm,height=7cm}}
\caption{ Figure 1(a), The hashed area represents 
the magnetic layers, and the clear areas represent normal regions. The 
black strip between the leads represent the portion of the variable-height  
barrier. Figure 1(b) shows the equivalent circuit of the structure shown in 1(a).}
\label{fig1}
\end{figure}

\begin{figure}
\centerline{\psfig{figure=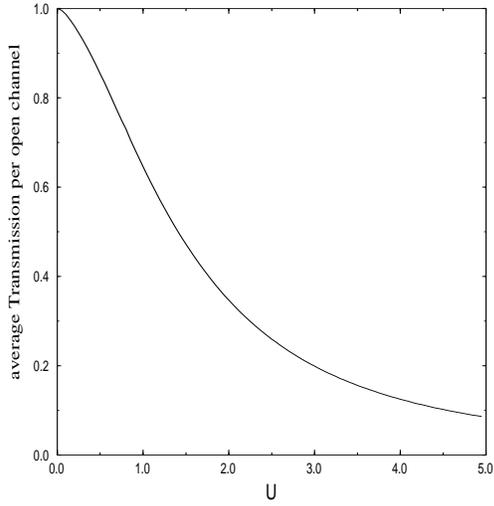,width=7cm,height=7cm}}
\caption{ The average transmission coefficient per open channel 
for a 2D tight-binding lattice. The on-site energy is $\epsilon_0$ and the
barrier strength $U$.
The barrier is 2 lattice sites wide, and the on-site 
energy is $\epsilon_0 + U$.}
\label{fig2}
\end{figure}

\begin{figure}
\centerline{\psfig{figure=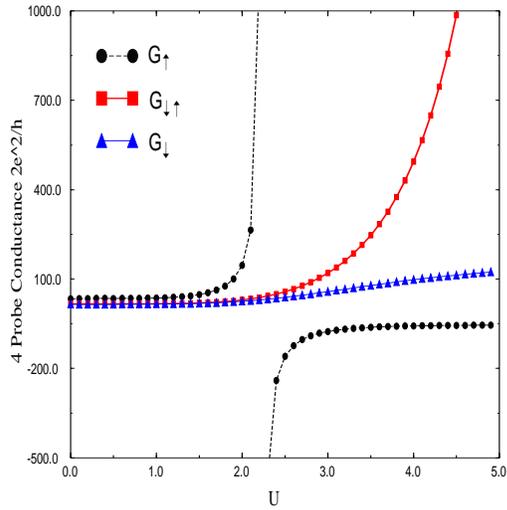,width=7cm,height=7cm}}
\caption{ The dependence of the 4 probe conductances 
 $G_{\uparrow}$,$G_{\uparrow\downarrow}$ 
and $G_{\downarrow}$ on barrier strength, for the 
structure shown in Fig.1.}
\label{fig4}
\end{figure}

\begin{figure}
\centerline{\psfig{figure=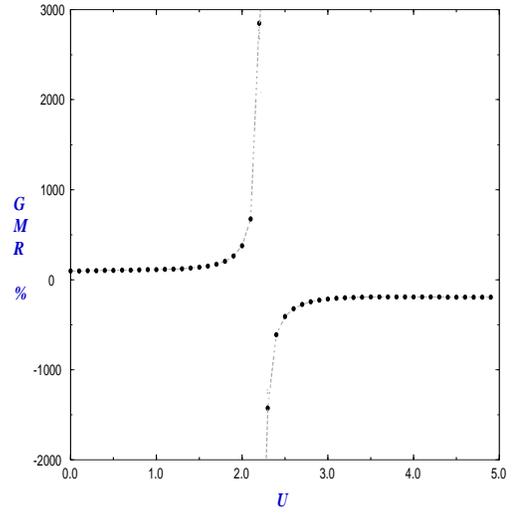,width=7cm,height=7cm}}
\caption{ The dependence of the GMR ratio 
on barrier strength. }
\label{fig3}
\end{figure}

\end{document}